\documentclass[epj,twocolumn]{webofc} 
\woctitle{ISVHECRI 2018} 
\usepackage[varg]{txfonts}

\begin{document}
\title{Measurements of very-forward energy with the CASTOR calorimeter of CMS}
\author{Sebastian Baur\inst{1} for the CMS Collaboration}
\institute{Institute for Nuclear Physics, Karlsruhe Institute of Technology (KIT)}

\abstract{The very-forward energy production in hadron collisions is of paramount importance for the understanding of ultra-high energy cosmic ray air showers. The CASTOR calorimeter of CMS is located at $-6.6<\eta<-5.2$ in the phase-space where the peak of energy is deposited at LHC. The composition and characteristics of the particles in this phase-space have a determining impact on the formation of air shower cascades. An overview of various energy measurements performed with CASTOR is reported and possible implications for cosmic ray physics are outlined.}

\maketitle

\section{Introduction}
\label{intro}
Particle production at forward rapidities in high energy hadronic collisions is to a large extent not accurately described by calculations from first principles. The relevant processes, especially multiparton interactions (MPI) and the fragmentation of the beam remnants, are modelled phenomenologically in Monte Carlo event generators with parameters tuned from data~\cite{Corke:2010yf,Skands:2014pea}.
A good understanding of forward particle production is important to accurately simulate the interactions of cosmic rays in the upper atmosphere as well as the subsequent development of extensive air showers~\cite{Ulrich}.
In addition, the production of forward charged hadrons has direct impact on the total number of air-shower muons at the ground, whose model predictions still show a deficiency compared to the data~\cite{Auger-Muons}.

\section{The CASTOR calorimeter}
The CASTOR calorimeter of the CMS experiment is a unique detector at the LHC and allows to perform dedicated measurements to address these questions. The CASTOR calorimeter is a sampling calorimeter composed of layers of fused silica quartz plates and tungsten absorbers. It is located only on the negative side of CMS and covers the pseudorapidity region $-6.6<\eta<-5.2$. The two front channels of each azimuthal segment (so called tower), have a combined depth of 20 radiation lengths and form the electromagnetic section of each tower. The energy recorded in these channels is dominated by energy deposits of electrons and photons, which include photons from neutral pion decays. The remaining 12 channels of each tower constitute the hadronic section. The full depth of a tower amounts to 10 hadronic interaction lengths. A more detailed description of the CMS detector can be found in Ref.~\cite{CMS-Detector}. A detailed description of the CASTOR calorimeter is given in Refs.~\cite{CASTOR-Prototype,CASTOR-Goettlicher}.

\section{Energy measurements with CASTOR}
\label{measurements}

CASTOR took data at many different centre-of-mass energies and beam configurations at the LHC. Here, we present highlighted measurements of the energy seen by CASTOR in proton-proton collisions during LHC Run 1 and Run 2 at $\sqrt{s}=0.9,\,2.76,\,7,\,\text{and }13\,$TeV. The data are compared to predictions of various hadronic interaction models, especially those relevant for the simulation of cosmic ray air showers, such as \textsc{epos}~\cite{EPOS,EPOS-LHC}, \textsc{QGSJetII}~\cite{QGSJet}, and \textsc{Sibyll}~\cite{Sibyll21,Sibyll23}.

\subsection{Underlying event at $\sqrt{s}=0.9,\,2.76,\,\text{and }7\,$TeV}

\begin{figure*}
\centering
\includegraphics[width=0.7\textwidth , trim={0 2cm 2cm 0}, clip]{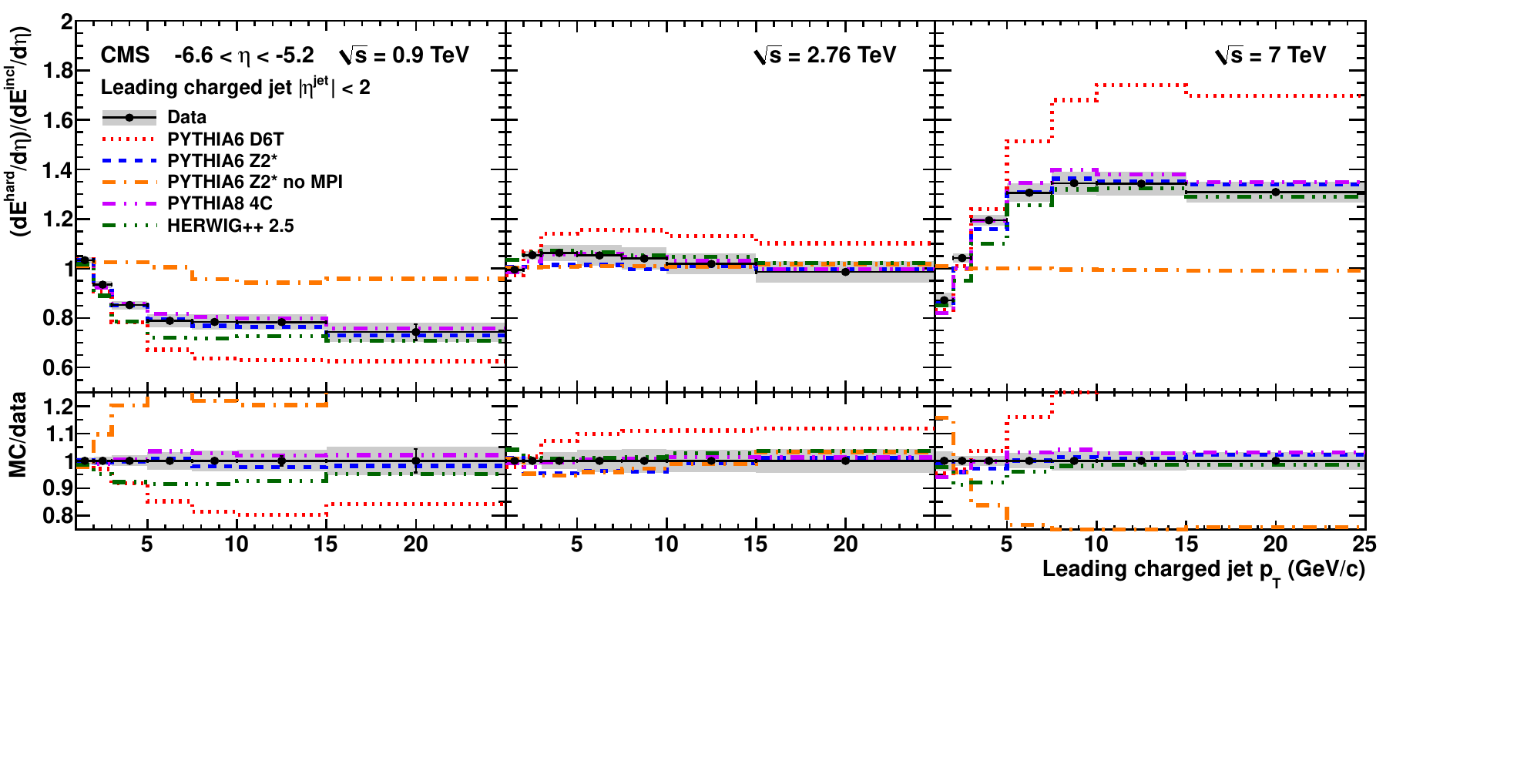}\\
\includegraphics[width=0.7\textwidth , trim={0 2cm 2cm 0}, clip]{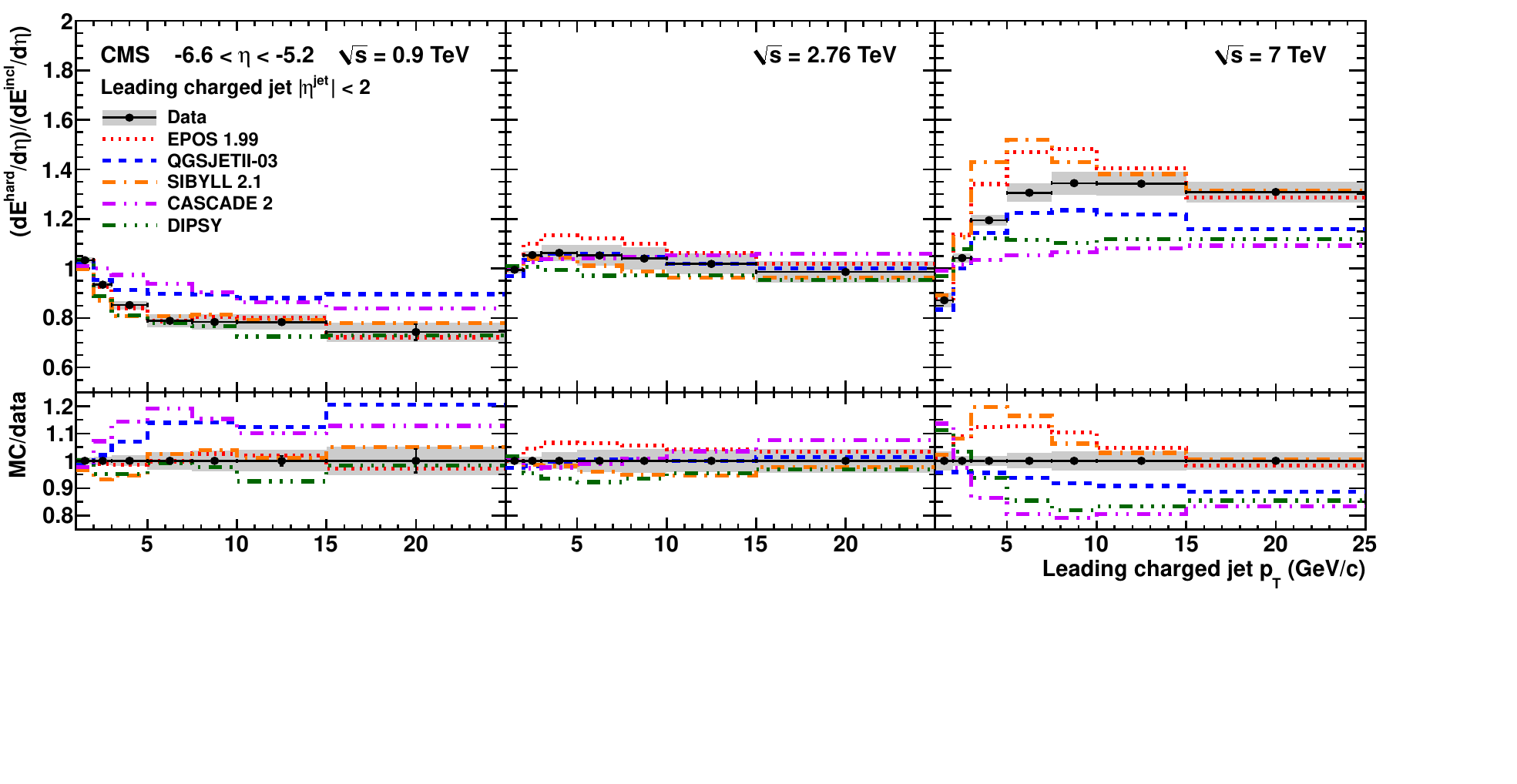}
\caption{Ratio of the energy deposited in the pseudorapidity range $-6.6<\eta<-5.2$ for events with a charged-particle jet with $|\eta_{\mathrm{jet}}|<2$ with respect to the energy in inclusive events, as a function of the jet transverse momentum $p_{\mathrm{T}}$ for $\sqrt{s}$ = 0.9 (left), 2.76 (middle), and 7 TeV (right). Data are compared to the \textsc{Pythia} and \textsc{herwig++} MC models (top) as well as to models used for air shower physics (bottom). Error bars indicate the statistical uncertainty on the data points, while the grey band represents the statistical and systematic uncertainties added in quadrature~\cite{CMS-UE-CASTOR-Run1}.}
\label{fig-1}
\end{figure*}

In Fig.~\ref{fig-1}, the ratio of the forward energy density, $\mathrm{d}E/\mathrm{d}\eta$, for events with a charged-particle jet produced at central pseudorapidity ($|\eta_{\text{jet}}| < 2$) relative to the forward energy density for inclusive events is shown. This forward energy density ratio is measured as a function of the central jet transverse momentum, $p_\mathrm{T}$, at three different centre-of-mass energies ($\sqrt{s} = 0.9,\,2.76,\,\text{and }7\,$TeV). The data are corrected for detector effects and compared to various different hadronic event generators~\cite{CMS-UE-CASTOR-Run1}.

It is observed that the evolution of the forward energy density behaves significantly different at different centre-of-mass energies. At $\sqrt{s}=0.9\,$GeV, the energy seen by CASTOR decreases as the central jet $p_\mathrm{T}$ increases. This is due to the fact that the energy is dominated by the fragmentation of the beam remnant and that more energy is taken from the remnant as the central activity increases. With increasing centre-of-mass energy, the contribution of MPI becomes more and more important to the energy in the CASTOR acceptance. The typical behaviour of the underlying event is seen at $\sqrt{s}=7\,$TeV, where a sharp rise, followed by a plateau region is observed. If MPI is turned off in \textsc{Pythia}6~\cite{pythia6}, the forward energy becomes independent of the central jet $p_\mathrm{T}$, which is ruled out by the data. Most studied event generators perform well in describing the general features of the data, although none describes all features at once.

\subsection{Forward energy density at $\sqrt{s}=13\,$TeV}
\begin{figure*}
\centering
\includegraphics[width=0.35\textwidth , clip]{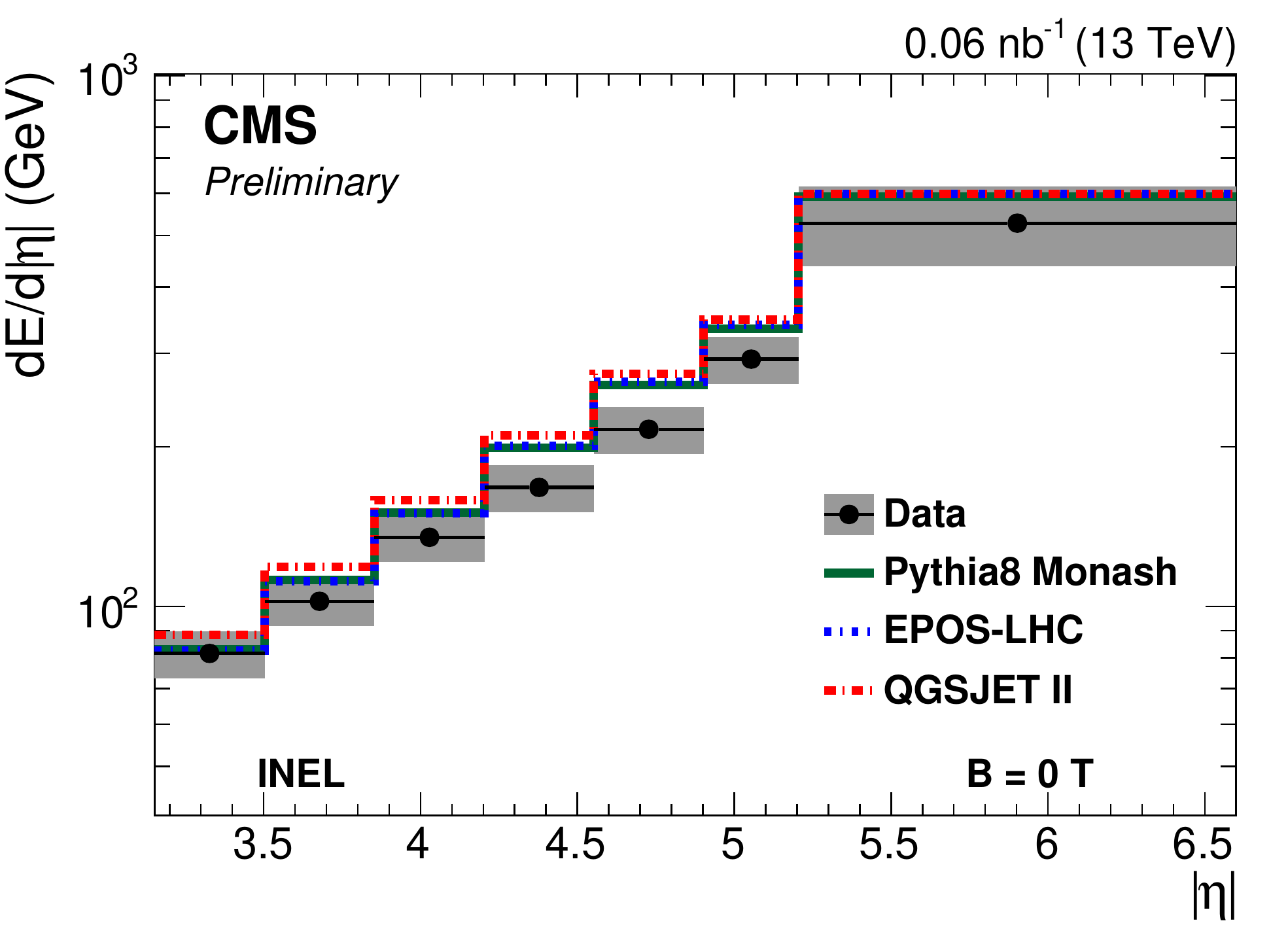}
\includegraphics[width=0.35\textwidth , clip]{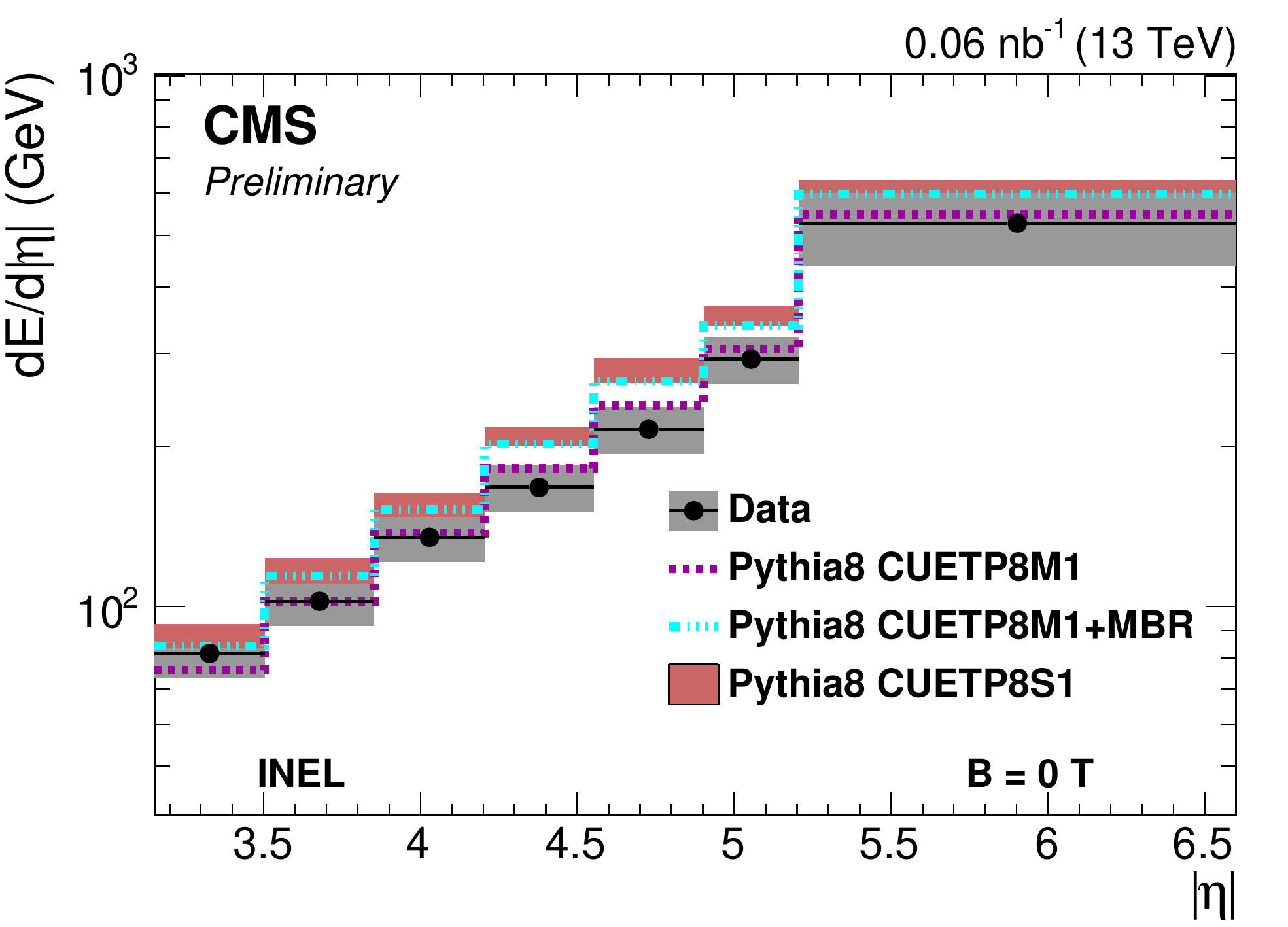}\\
\includegraphics[width=0.35\textwidth , clip]{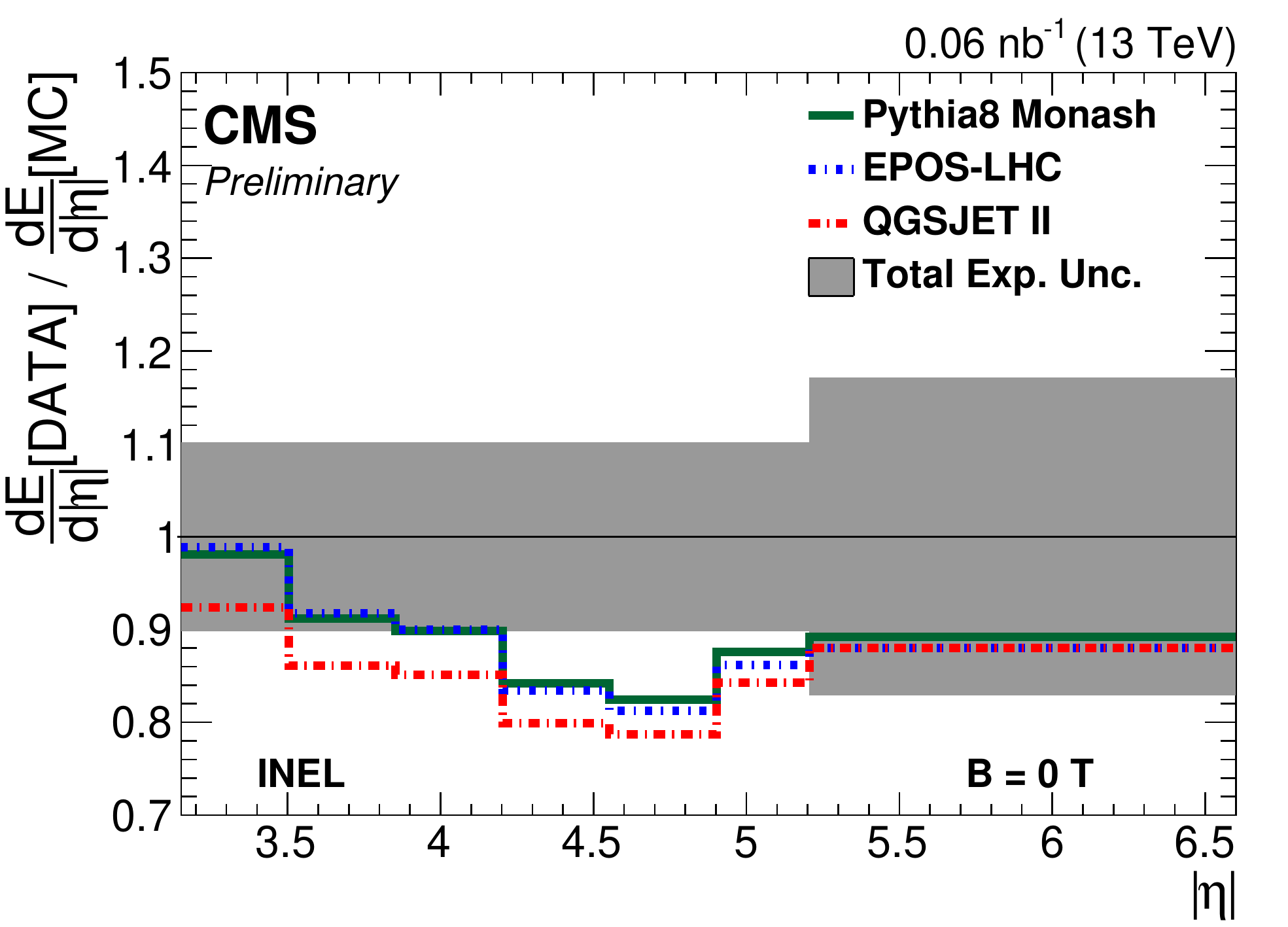}
\includegraphics[width=0.35\textwidth , clip]{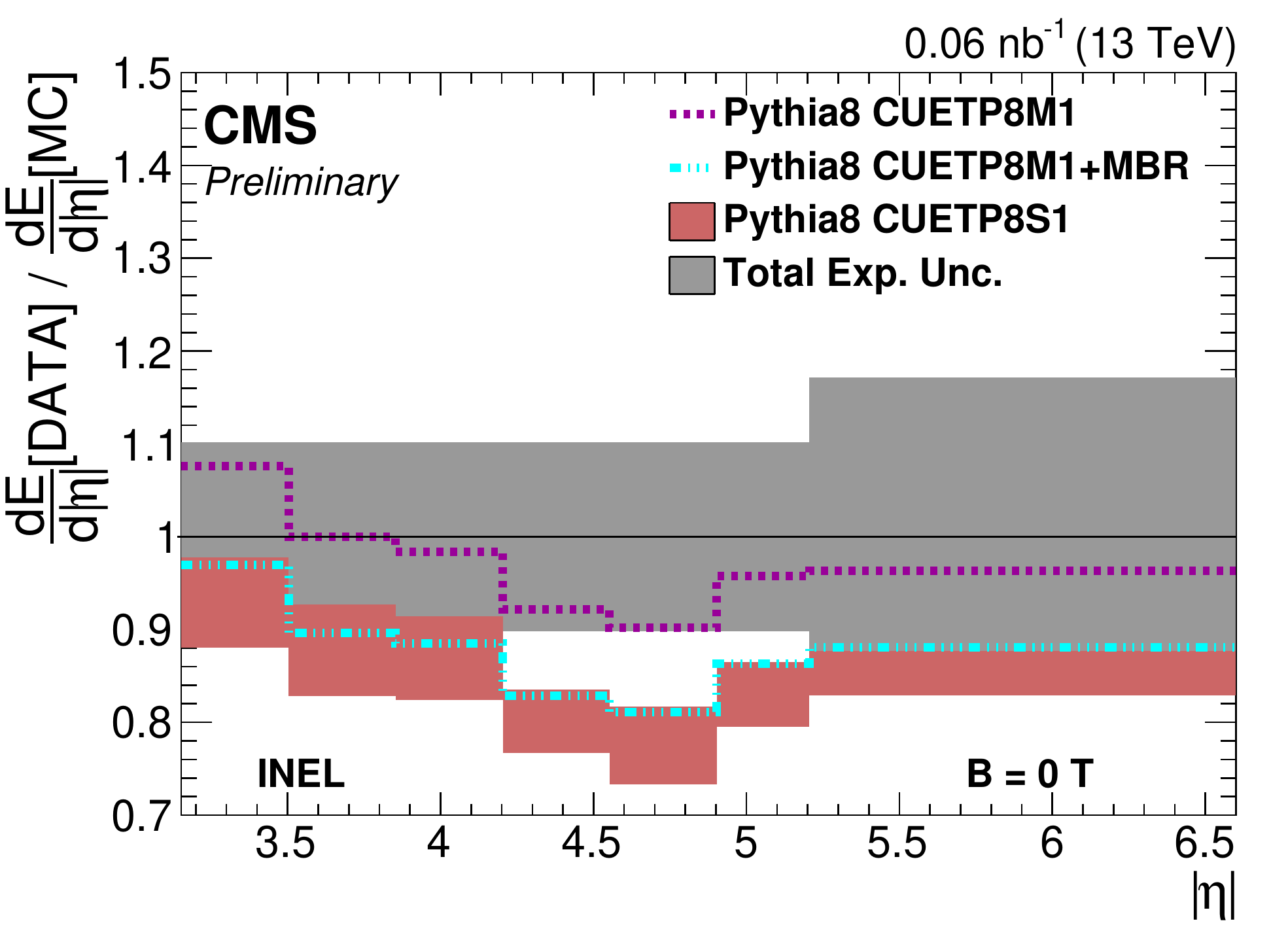}
\caption{Energy density at particle level as a function of pseudorapidity at 13 TeV for inclusive-inelastic events compared to predictions of various event generators. The grey band shows the total systematic uncertainty correlated across pseudorapidity bins. The bottom panel shows the ratio of the data to model predictions~\cite{CMS-EFlow-Run2}.}
\label{fig-2}
\end{figure*}

At $\sqrt{s}=13\,$TeV, the forward energy density is measured as a function of the pseudorapidity using CASTOR and the Hadron Forward (HF) calorimeter of CMS in the range $3.15<|\eta|<6.6$~\cite{CMS-EFlow-Run2}. The energy density is measured for different classes of inelastic collisions. Fig.~\ref{fig-2} shows the energy density for the inclusive-inelastic event selection, which requires a minimal fractional proton momentum loss $\xi>10^{-6}$, see Ref.\cite{CMS-EFlow-Run2} for details.

The considered models provide a reasonable description of the measured energy flow. The best description of the data is provided by the \textsc{Pythia}8 tune CUETP8M1~\cite{Pythia8-Tunes}. However, the evolution of the energy with $\eta$ is not well described by all models, especially in the region of the HF coverage at $3.15<|\eta|<5.20$. The spread of the model predictions is larger than the tuning uncertainties, illustrated by a red band given for \textsc{Pythia}8 CUETP8S1, thus inherent model differences are resolved.

\subsection{Forward energy spectra at $\sqrt{s}=13\,$TeV}
\begin{figure*}
\centering
\includegraphics[width=0.35\textwidth , clip]{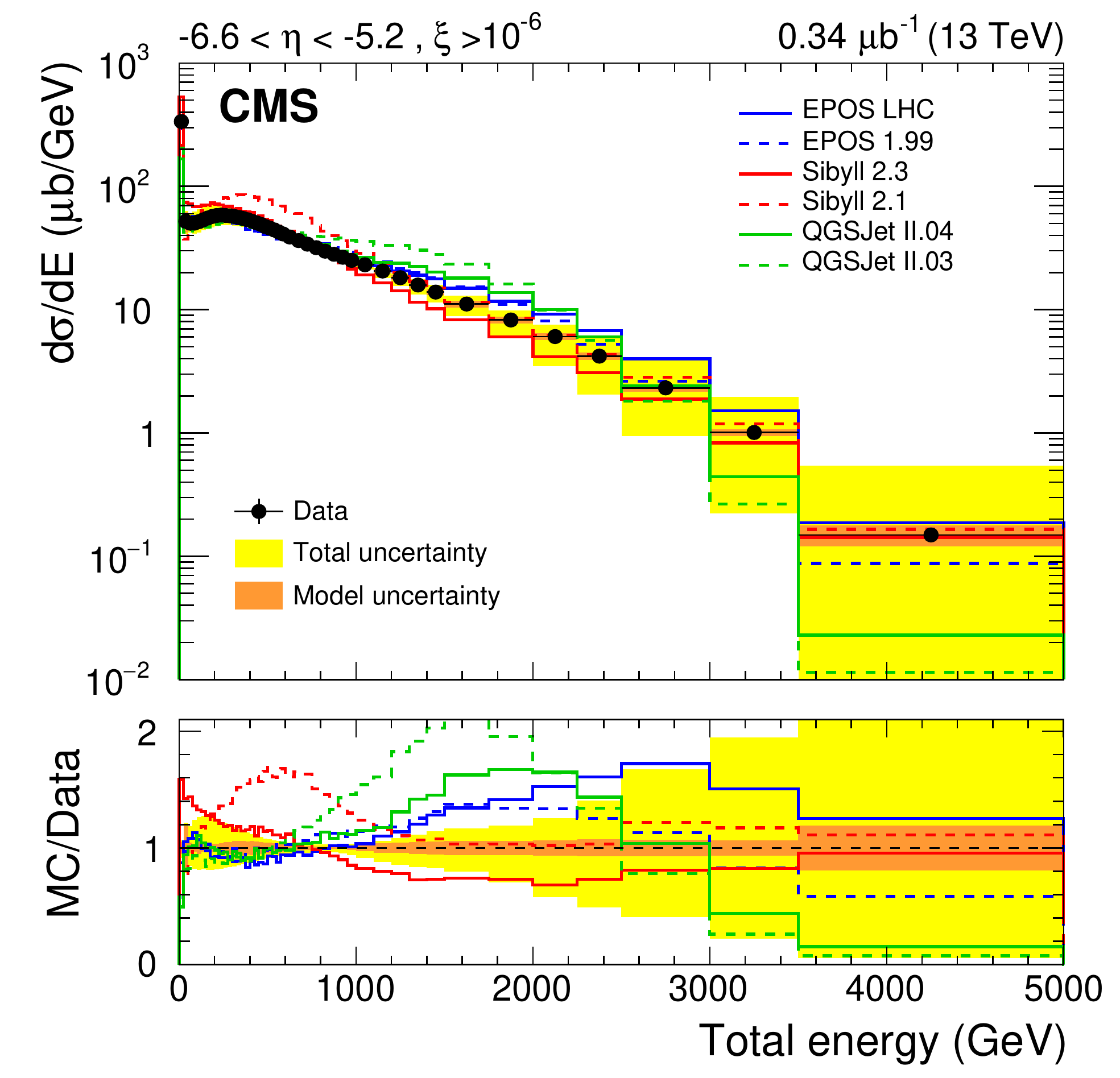}
\includegraphics[width=0.35\textwidth , clip]{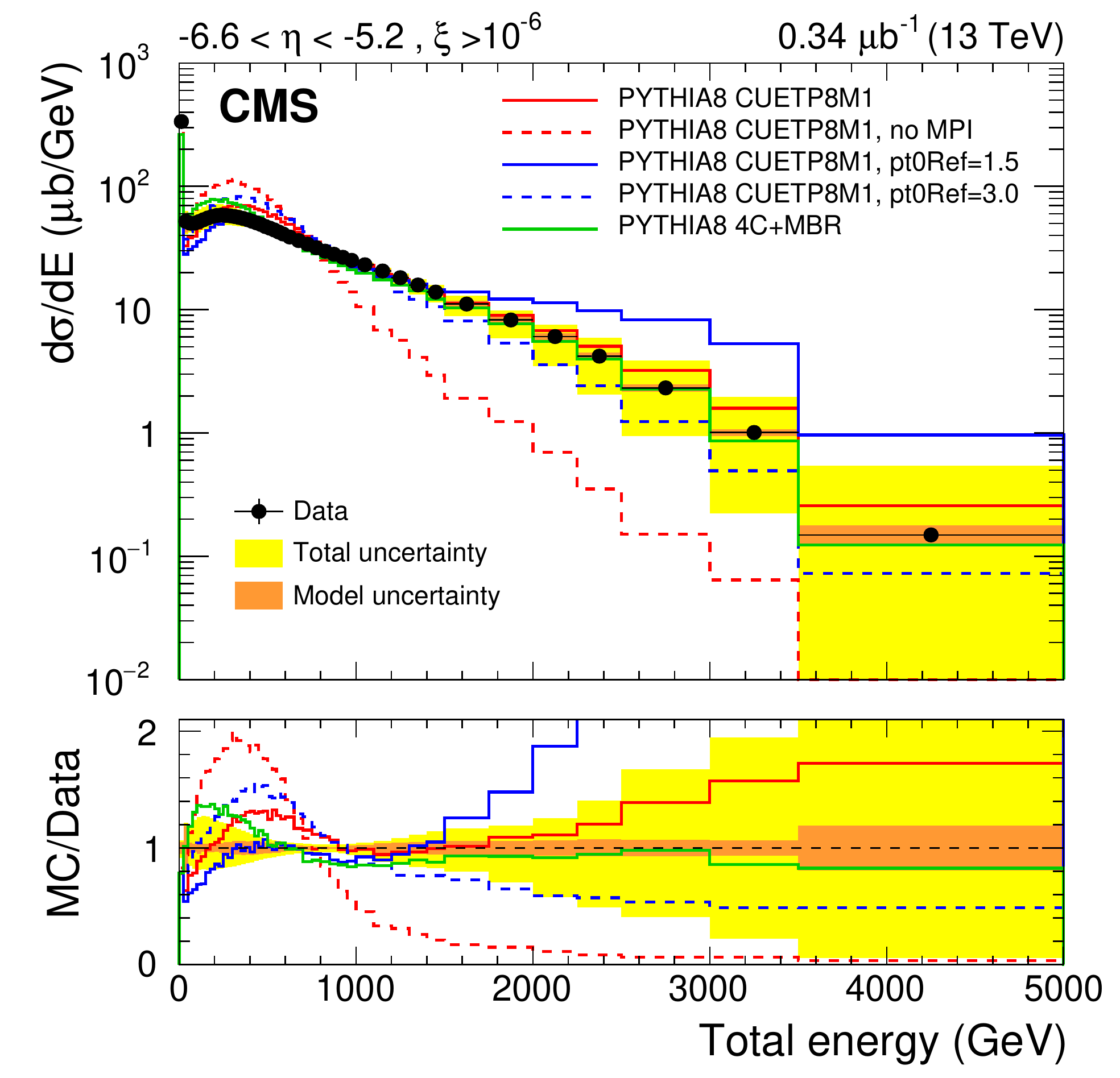}\\
\includegraphics[width=0.35\textwidth , clip]{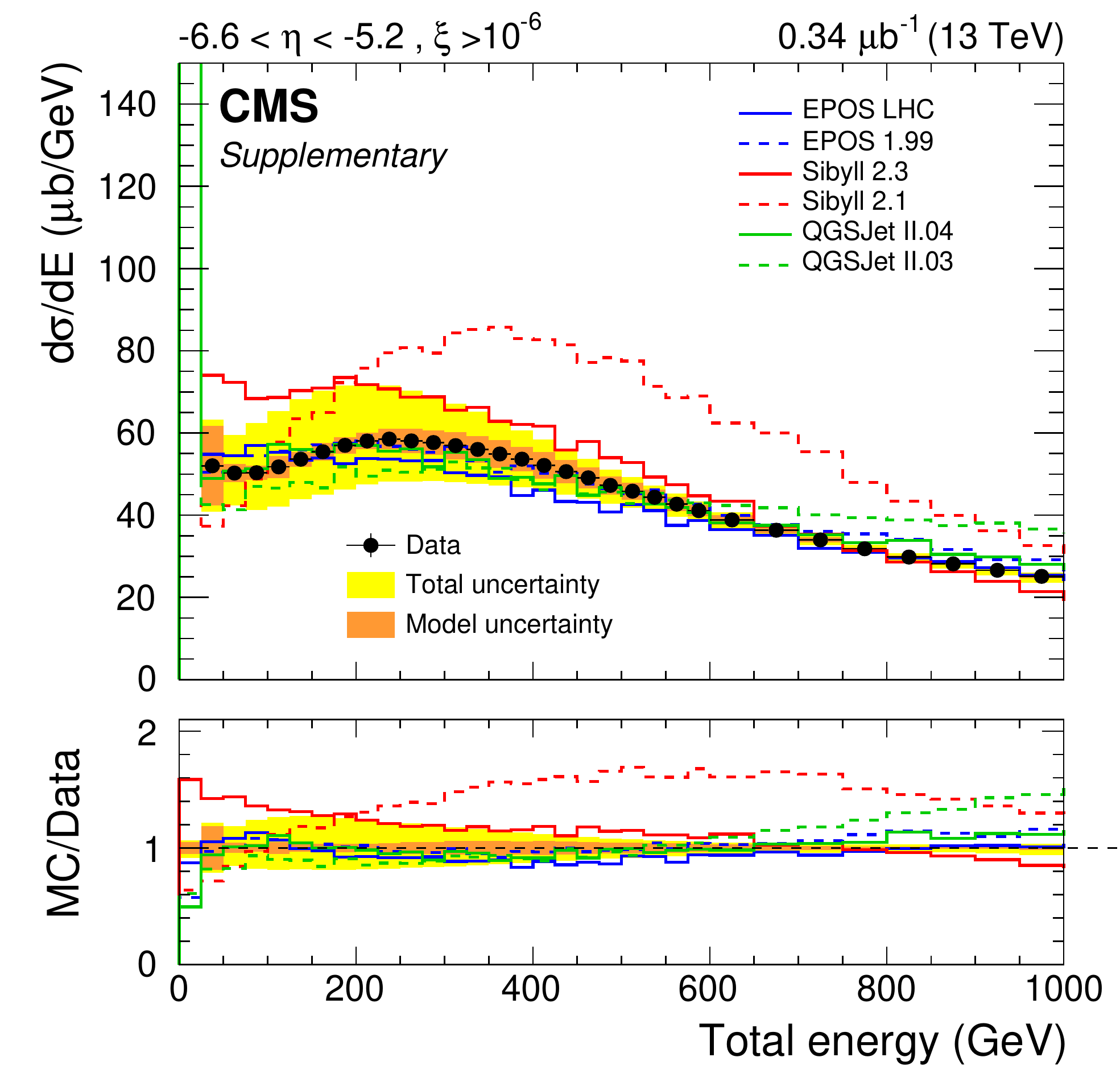}
\includegraphics[width=0.35\textwidth , clip]{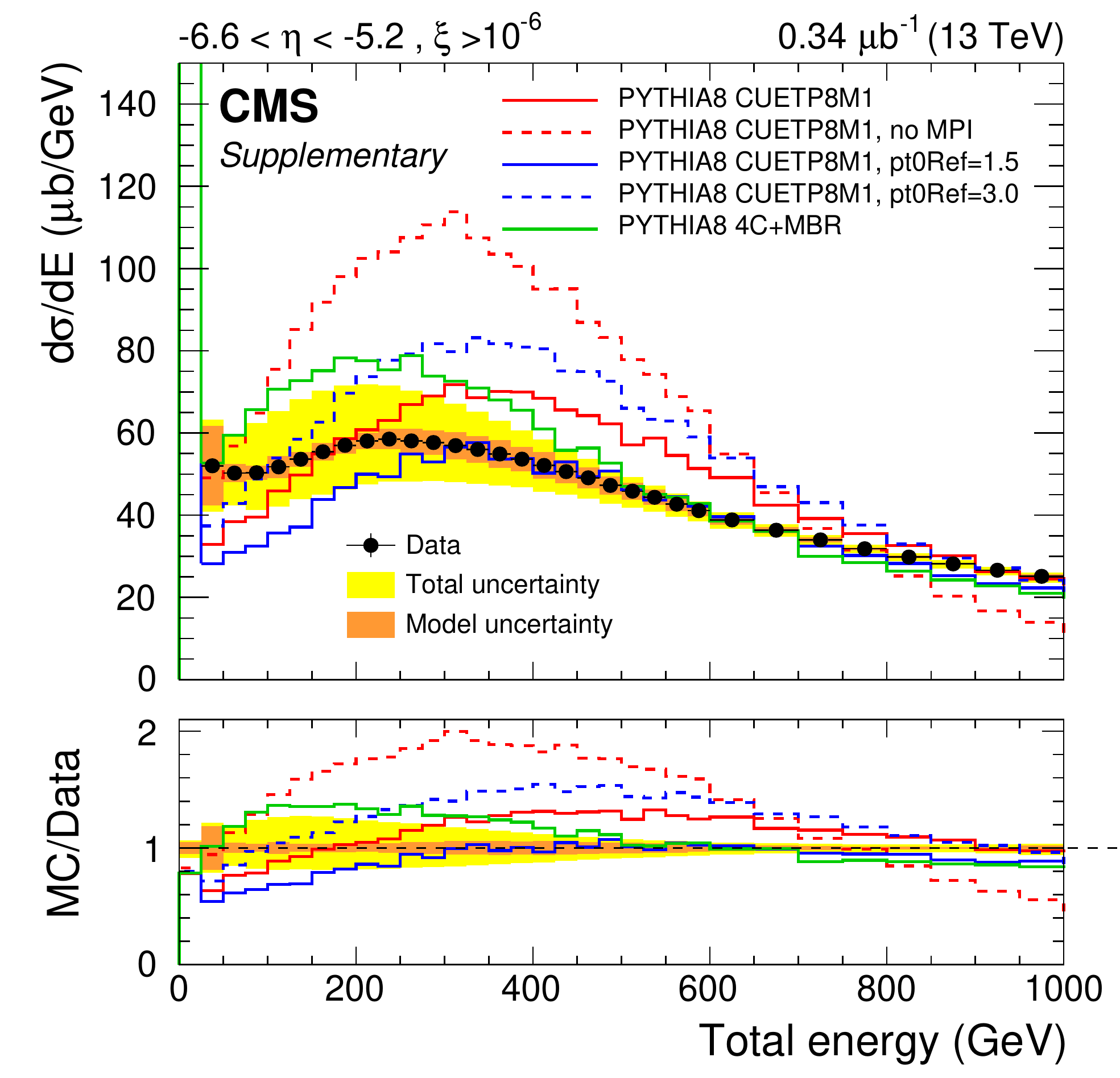}
\caption{Differential cross section as a function of the total energy in $-6.6<\eta<-5.2$. The left panel shows the data compared to MC event generators mostly developed for cosmic ray induced air showers, and the right panel to different \textsc{Pythia}8 tunes. The bottom panel shows the same data with linear scale at low energies. The yellow band indicates the total uncertainty of the measurement, the orange band the model-uncertainty due to the unfolding~\cite{CMS-CASTOR-Spectra-Run2}.}
\label{fig-3}
\end{figure*}

\begin{figure*}
\centering
\includegraphics[width=0.35\textwidth , clip]{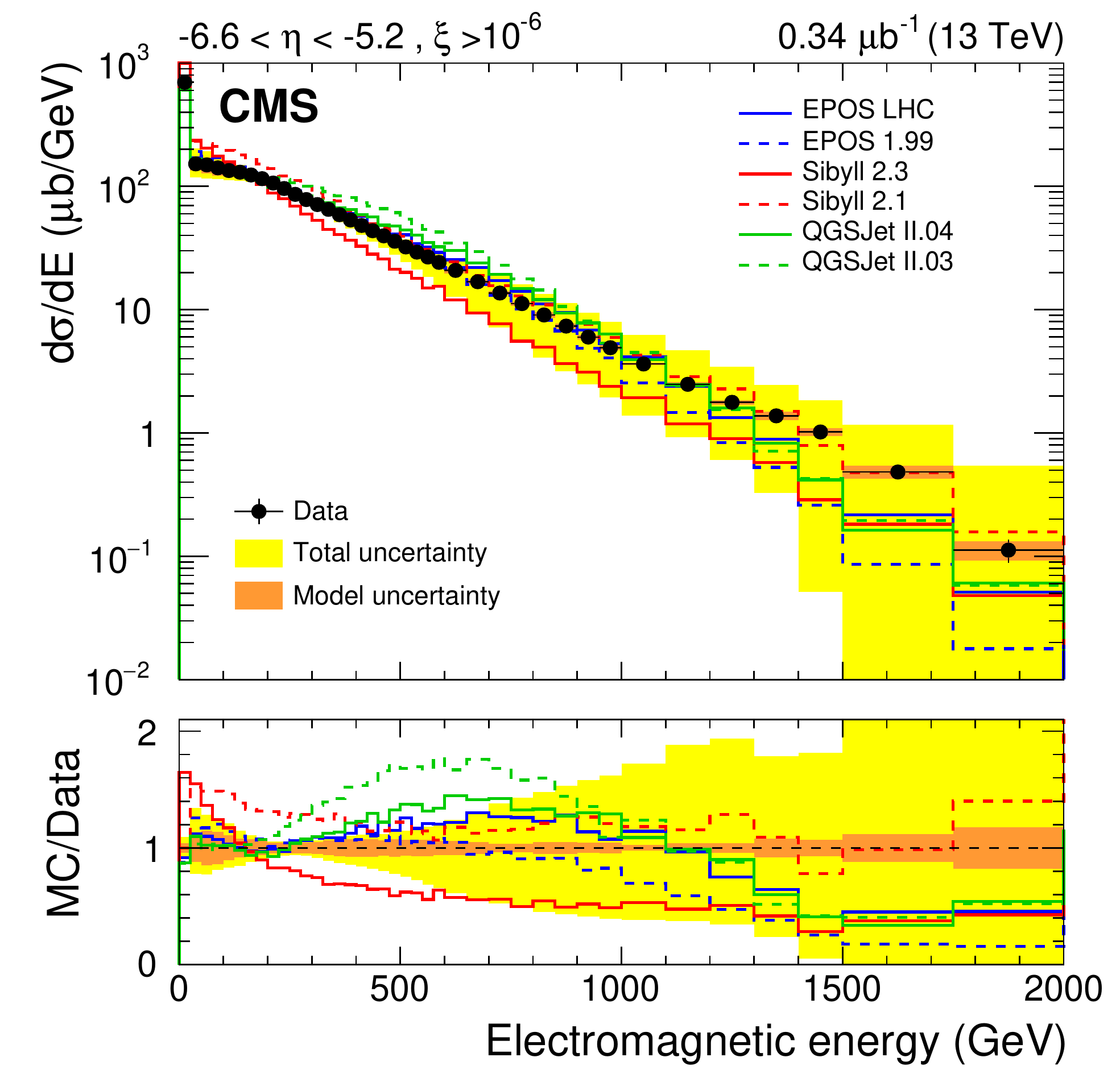}
\includegraphics[width=0.35\textwidth , clip]{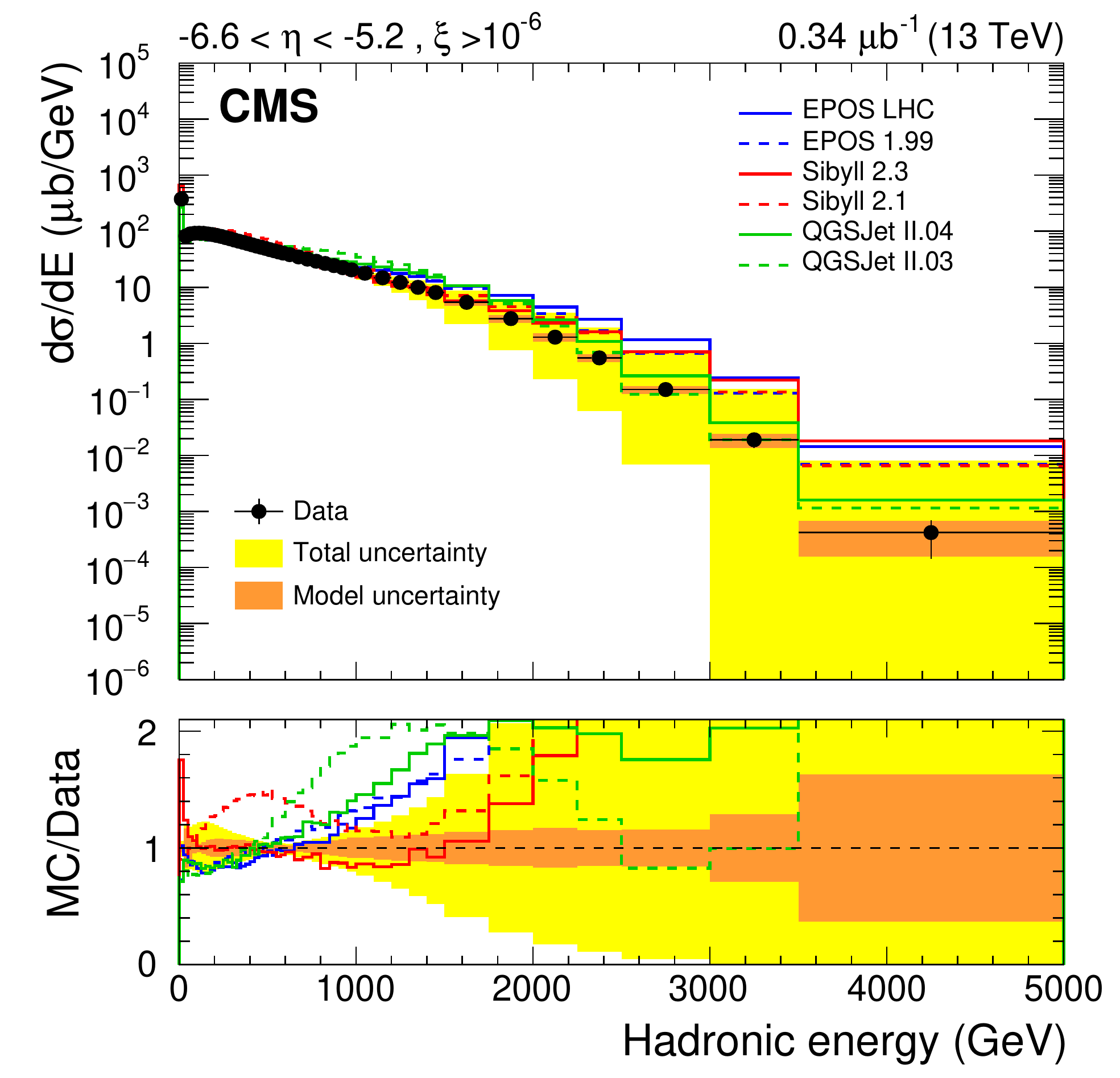}\\
\caption{Differential cross section as a function of the electromagnetic energy (left) and the hadronic energy (right). The yellow band indicates the total uncertainty of the measurement, the orange band the model-uncertainty due to the unfolding~\cite{CMS-CASTOR-Spectra-Run2}.}
\label{fig-4}
\end{figure*}

For the same class of events with $\xi>10^{-6}$, the energy distribution within the CASTOR acceptance is studied in more detail. The differential cross section as a function of energy in proton-proton collisions at a centre-of-mass energy of 13\,TeV is shown in Figs.~\ref{fig-3} and \ref{fig-4}. Here for the first time, the possibility of CASTOR to separate electromagnetic and hadronic energy depositions is used~\cite{CMS-CASTOR-Spectra-Run2}. Therefore, the measurement is performed as a function of the total energy deposited in CASTOR (Fig.~\ref{fig-3}), as well as of the electromagnetic (energy of electrons and photons only) and hadronic (energy of charged and neutral hadrons) components (Fig.\ref{fig-4}).

The spectra are sensitive to the modelling of multiparton interactions, which is demonstrated by changes of the parameter $p_{\mathrm{T,0}}^{\mathrm{ref}}$ of \textsc{Pythia}8. Furthermore, the overall collision elasticity and the amount of diffraction influences the spectra shape at low energies. It can, for example, be observed that the fraction of events with little energy is significantly overestimated by \textsc{Sibyll 2.3} compared to the data. This can be a hint of a too large elasticity in the model. The separation into the electromagnetic and hadronic contribution provides additional constraints for the models. While the electromagnetic energy spectrum is well described by the latest generation of interaction models, the hadronic energy is slightly overestimated at the edge of the uncertainties. The data suggest that there is therefore no room to increase the number of air shower muons by increasing the hadronic energy in proton-proton collisions.

\section{Summary}
The CASTOR calorimeter of CMS is a unique detector in the forward phase space. A series of measurements was performed to study the energy density and distribution in this acceptance in order to perform benchmark tests on hadronic event generators with the special goal to improve the understanding of the development of extensive air showers.

Relative energy density as function of the central jet $p_\mathrm{T}$ probes the transition between the remnant fragmentation and MPI dominated regime as the centre-of-mass energy increases. The forward energy density as a function of pseudorapidity and the inclusive energy spectra in the CASTOR acceptance have good sensitivity to the modelling of multiparton interactions and are furthermore sensitive to diffraction. In addition, the first measurement separating the electromagnetic and hadronic energy in the same phase-space was performed. This constrains the modelling of muon production in extensive air showers.

\end{document}